\documentclass[dvips,eps,pra,tightenlines,twocolumn,showpacs]{revtex4}

\usepackage{graphicx}
\usepackage{bbm}
\usepackage{amsmath}
\usepackage{amsfonts}
\usepackage{amssymb}
\usepackage{hhline}
\usepackage{theorem}
\theoremstyle{plain}

\begin{document}
\title{Multipartite entanglement measure for all discrete systems}
\author{Beatrix C. Hiesmayr and Marcus Huber\\
    \small{Faculty of Physics, University of Vienna,
    Boltzmanngasse 5, A-1090 Vienna, Austria}}

\begin{abstract}
Via a multidimensional complementarity relation we derive a novel
operational entanglement measure for any discrete quantum system,
i.e. for any multidimensional and multipartite system. This new
measure admits a separation into different classes of entanglement
obtained by using a flip operator $2$--, $3$--,\dots, $n$--times,
defining a $m$--flip concurrence. For mixed states bounds on this
$m$--flip concurrence can be obtained. Moreover, the information
content of a $n$--partite multidimensional system admits a simple
and intuitive interpretation. Explicitly, the three qubits system is
analyzed and e.g. the physical difference in entanglement of the
$W$--state, the $GHZ$--state or a bi--separable state is revealed.
%
%We introduce a novel approach to derive a multidimensional
%complementarity relation. From this we derive a novel operational
%entanglement measure for any discrete quantum system, i.e. for any
%multidimensional and multipartite system. This new measure admits an
%separation into different classes of entanglement obtained by using
%a flip operator $2$--, $3$--,\dots, $n$--times, defining a $m$--flip
%concurrence. For mixed states bounds on this $m$--flip concurrence
%can be obtained. Moreover, the information content of a $n$--partite
%multidimensional system can be viewed as  a part containing only
%information obtainable by the subsystems and into a part containing
%entanglement.
\end{abstract}
\maketitle

\section{Introduction}
For many quantum mechanical applications entanglement is the basic
ingredient. Mathematically entanglement is well defined. However, no
simple operational criterion to detect entanglement versus
separability is known. Especially with multipartite entanglement, which is
subject to recent research (Ref.~\cite{theory3,theory4,theory5,theory6,theory7,theory8,Meyer,Boixo}),
there are still many open questions regarding its properties.
Moreover, it is known that there exist different ``kinds'' of entanglement, see e.g. Ref.~\cite{Horodeckibig}.
It is therefore highly desirable to first find an operational measure of entanglement and
secondly it should provide a good classification of the different
kinds of entanglement as this is the physical property which is
explored by various applications such as e.g. quantum cryptography
or quantum communication, see also Ref.~\cite{theory1,experiment1,experiment2}.

In this paper we provide both by defining a novel and very intuitive
entanglement measure which is additive for pure states and has the
advantage to separate entanglement into $2$--, $3$--,\dots,
$n$--flip entanglement and for certain cases even into bipartite,
tripartite,\dots $n$--partite entanglement. It works for any
dimension and any number of particles.

\begin{figure}
(a)\includegraphics[width=3cm,keepaspectratio=true]{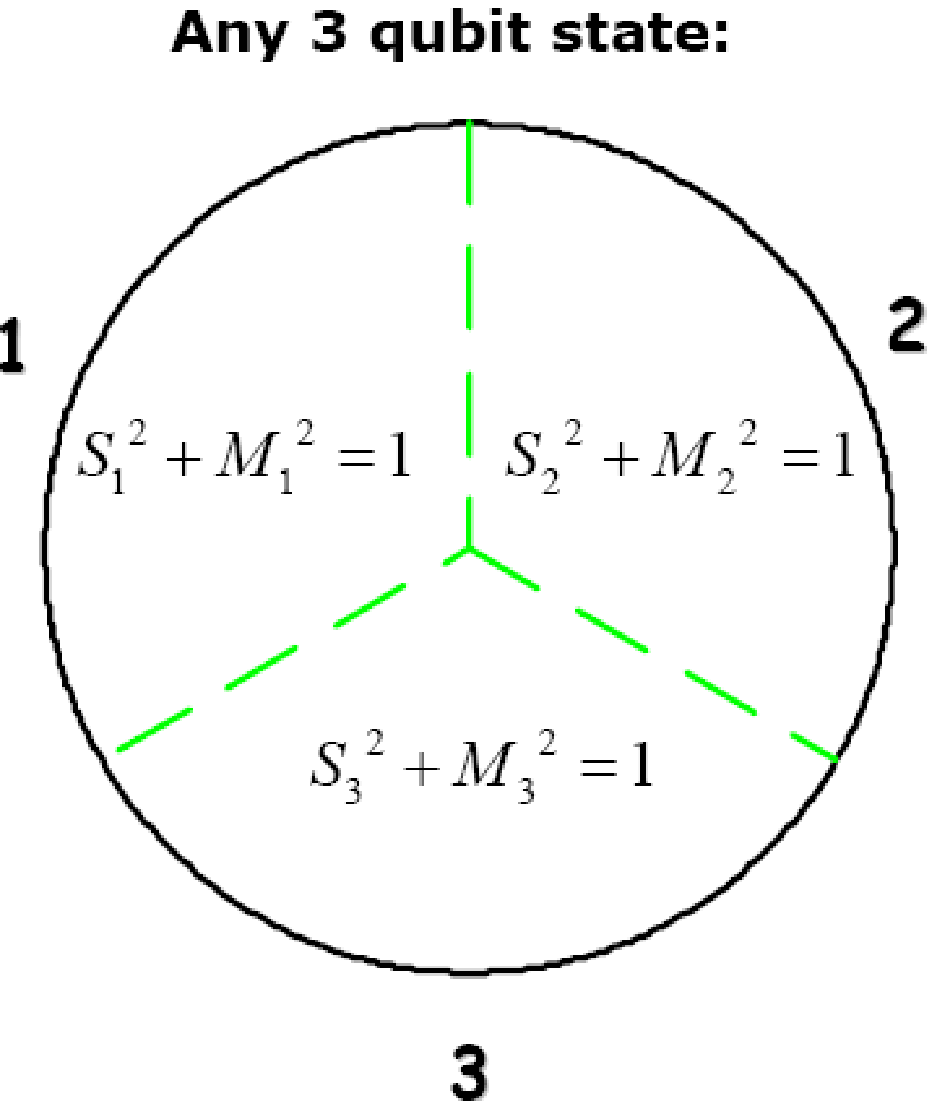}
(b)\includegraphics[width=3cm,keepaspectratio=true]{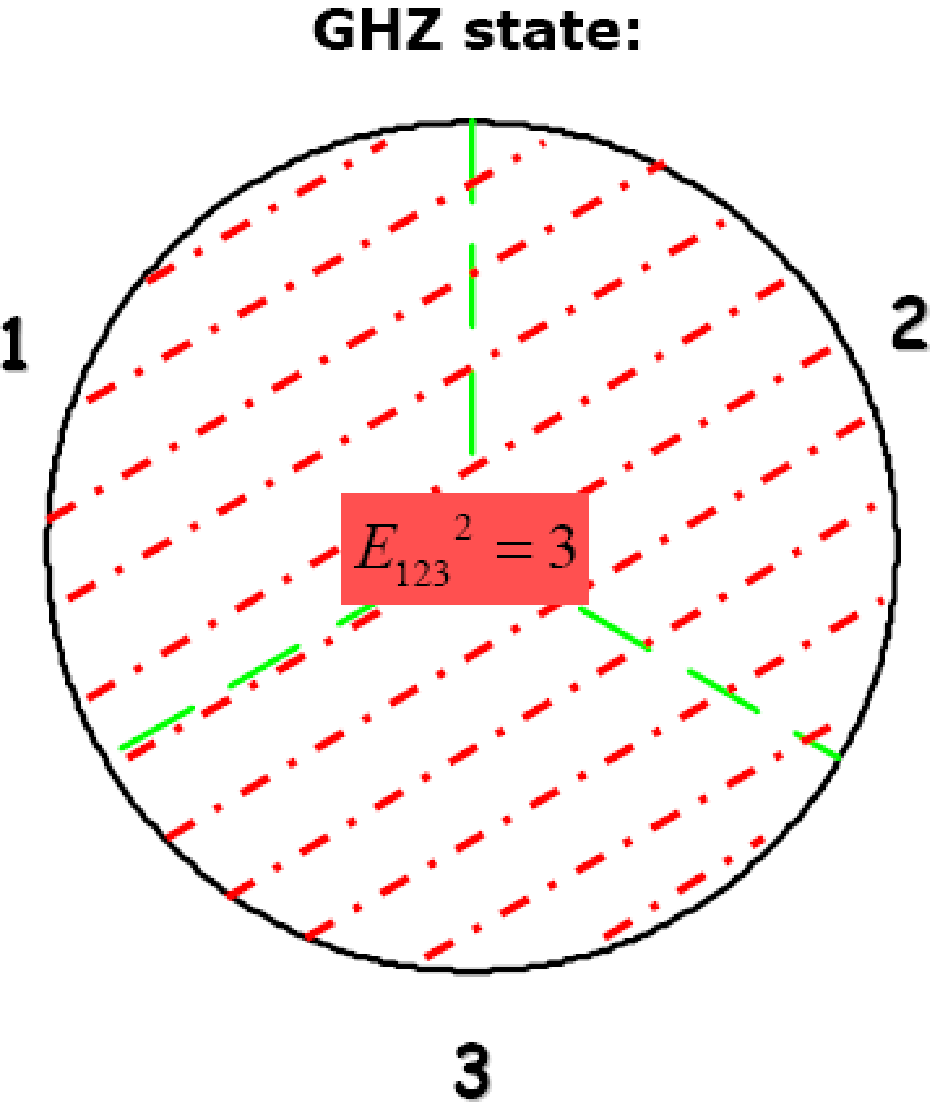}
(c)\includegraphics[width=3cm,keepaspectratio=true]{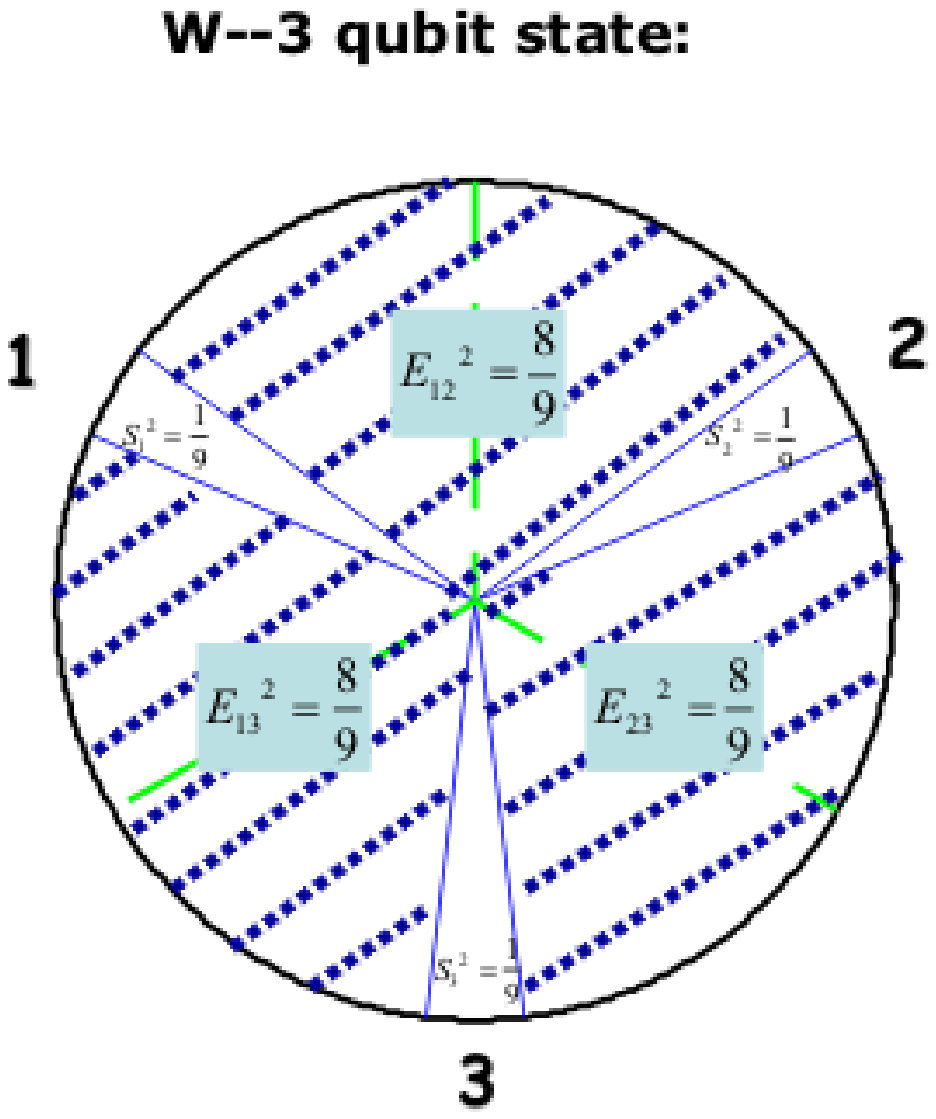}
(d)\includegraphics[width=3cm,keepaspectratio=true]{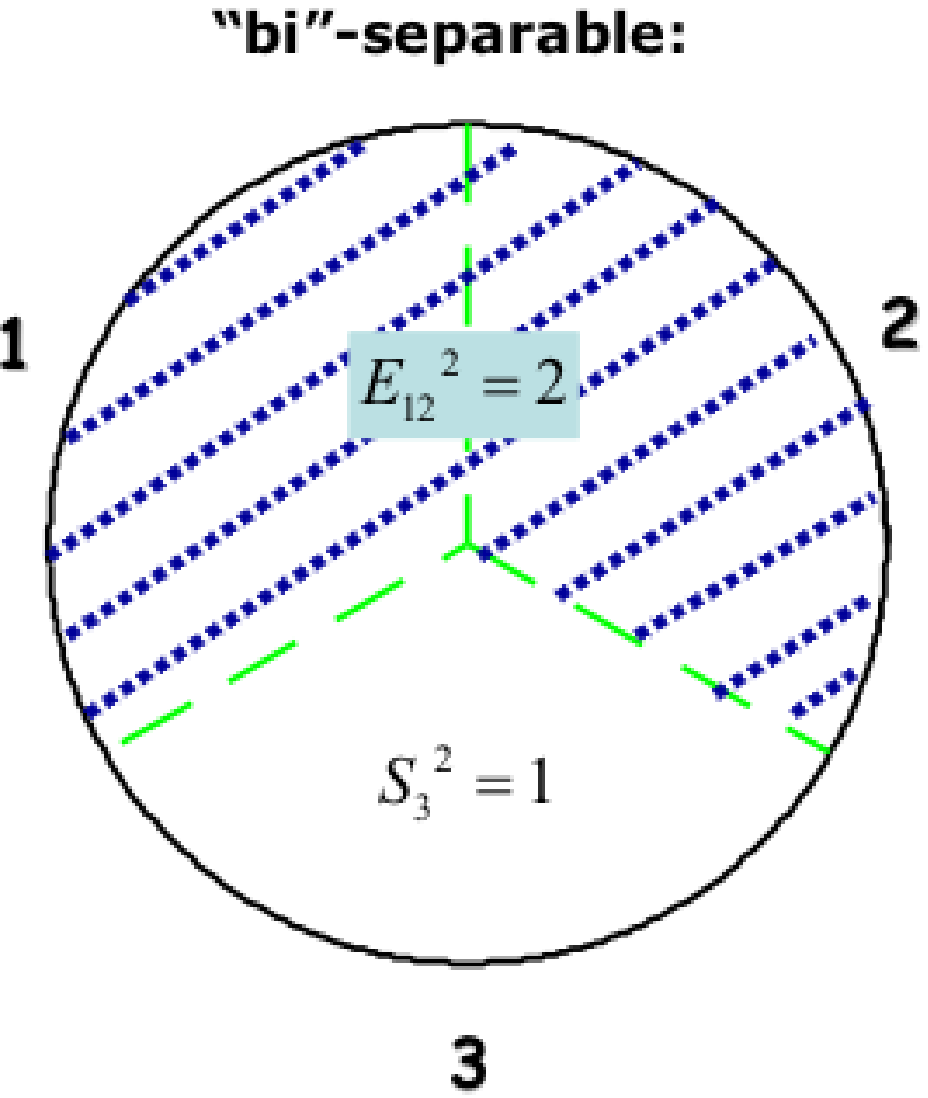}
\caption{Here the information content of three qubits states,
$3=\sum_{s=1}^3 (S^2_s+M_s^2)=\sum_{s=1}^3
S^2_s+E_{12}+E_{23}+E_{13}+E_{123}$, is visualized. In (a)
Bohr's complementarity relation for each qubit is drawn. (b)
visualizes the GHZ--state,
$\frac{1}{\sqrt{2}}\lbrace|000\rangle+|111\rangle\rbrace$, which is
a genuine tripartite state, the bipartite and single properties are
zero. In (c) the W--state,
$\frac{1}{\sqrt{3}}\lbrace|001\rangle+|010\rangle+|100\rangle\rbrace$,
is visualized, the single properties are nonzero as well as the
bipartite entanglement. In (d) the ``bi''--separable state,
Eq.~(\ref{halvesep}), is drawn, which shows
as desired only bipartite entanglement though the $3$--flip
concurrence is nonzero. Therefore, as desired, the particle $3$ is
independent of particle $1$ and $2$.}\label{fig:anystate}
\end{figure}

Explicitly, we show for three qubit systems how this novel measure
admits the separation of entanglement into genuine bipartite and
tripartite entanglement. And moreover how its substructure is
revealed, i.e. the entanglement property of each qubit with all
others, see also Fig.~\ref{fig:anystate}.

For that we start by proposing the following separation of the
information content in a $n$--partite quantum state of arbitrary
dimension $\rho$:
\begin{eqnarray}\label{voll}
\underbrace{I(\rho)+R(\rho)}_{\textrm{single
property}}+\underbrace{E(\rho)}_{\textrm{entanglement}}=n
\end{eqnarray}
where
\begin{eqnarray}
I(\rho)&:=&\sum_{s=1}^n\underbrace{{\cal
S}_s^2(\rho)}_{\textrm{single property of subsystem s}}
\end{eqnarray}
 We will show that for certain cases:
\begin{itemize}
    \item $I(\rho)$ contains all locally obtainable information,
    \item $E(\rho)$ contains all information encoded in
    entanglement,
    \item $R(\rho)$ is the complementing missing information,
     due to a classical lack of knowledge about the quantum state.
\end{itemize}
Moreover, we show how the total amount of entanglement can be
separated into m-flip concurrences:
\begin{eqnarray}
E(\rho)&:=&\underbrace{\textbf{C}_{(2)}^2(\rho)}_{\textrm{two flip
concurrence}}+\underbrace{\textbf{C}_{(3)}^2(\rho)}_{\textrm{three
flip concurrence}}+\;\quad(\dots)\nonumber\\&&+\underbrace{\textbf{
C}_{(n)}^2(\rho)}_{\textrm{n-flip concurrence}}\,.
\end{eqnarray}
Furthermore we show that with help of the m-flip concurrence we can,
at least for three qubits and possibly for even more complex
systems, indeed find a quantity interpretable as:
\begin{eqnarray}
E(\rho)&=&\underbrace{E_{(2)}(\rho)}_{\textrm{bipartite
entanglement}}+\underbrace{E_{(3)}(\rho)}_{\textrm{tripartite
entanglement}}
\end{eqnarray}
with the substructure:
\begin{eqnarray}
\underbrace{E_{(2)}(\rho)}&=&E_{(12)}(\rho)+E_{(23)}(\rho)+E_{(13)}(\rho)
\end{eqnarray}
We proceed in defining or deriving step by step the involved
quantities and discuss their physical content.% and di the two equations (\ref{voll}) and (\ref{part}).

\section{The single property ${\cal S}_s$ and Bohr's complementary
relation:} Bohr's complementary relation was first discussed to
understand the double slit experiment, its information theoretic
content can be formulated by the following formula
\cite{GreenbergerYasin,Englert}:
\begin{eqnarray}\label{comp}
{\cal S}^2(\rho):={\cal P}^2(\rho)+{\cal
C}_{\textrm{coh}}^2(\rho)\leq 1\;,
\end{eqnarray}
where  for all pure states the equality sign is valid. ${\cal
C}_{coh}$ is the coherence or in the case of the double slit
scenario the fringe visibility which quantifies the sharpness or
contrast of the interference pattern (``the wave--like property'').
Whereas ${\cal P}$ denotes the path predictability, i.e., the
\textit{a priori} knowledge one can have on the path taken by the
interfering system (``the particle--like property''). In double slit
experiment it is simply defined by ${\cal P}\;=\;|P_{I}-P_{II}|$,
where $P_I$ and $P_{II}$ are the probabilities for taking each path
($P_I+P_{II}=1)$.

As has been shown this complementary relation is useful to
understand several interfering two state system as e.g.
particle--antiparticle mixing systems~\cite{HH1,SBGH3} or Mott
scattering experiments of identical nuclei \cite{SBGH3} or even
specific thermodynamical quantum system \cite{HV}.

One can make Bohr's complementary relation always exact by adding
the quantity (dimensionality  $d$, here $2$)
\begin{eqnarray}\label{qubitmixedness}
M^2(\rho)=\frac{d}{d-1}\left(1-\, Tr(\rho^2)\right)
\end{eqnarray}
to the single particle property ${\cal S}$
\begin{eqnarray}\label{compmixed}
{\cal S}^2(\rho)+ M^2(\rho)= 1
\end{eqnarray}
for all states (pure: $M(\rho)=0$). $M(\rho)$ measures the mixedness
or linear entropy which equals in this case the uncertainty of
individual particles under investigation, clearly a ``classical''
uncertainty.

The complementarity principle seems to be an intrinsic property of
all discrete quantum systems. So even considering various dynamics
that a quantum system can be exposed to \cite{HH1}, the two
dimensional complementarity relation still holds true. The next
logical step is trying to generalize this relation to for a qudit
system.

\section{Bohr's complementary relation for $d$--dimensional
systems}: For a qudit system or a multi--slit system the definition of
predictability is not straightforward. One approach has recently
been introduced in Ref.~\cite{Janos2,Bergou}, however, we introduce
a similar approach which makes a generalization to multipartite
systems possible. To do that we will first introduce the following
useful quantity
\begin{eqnarray}
P_{i,j}&:=&Tr\left(\rho\,|i\rangle\langle j|\right)
\end{eqnarray}
and propose the generalized predictability for a qudit state $\rho$
\begin{eqnarray}
\lefteqn{P_g(\rho):=}\nonumber\\
&&\sqrt{\frac{d-1}{d}\sum_\pi\left|P_{0,0}-\frac{P_{1,1}+
P_{2,2}+(...)+P_{d-1,d-1}}{d-1}\right|^2}\nonumber\\
&=&\sqrt{\frac{d}{d-1}\sum_i |P_{i,i}-\frac{1}{d}|^2}=
\sqrt{\frac{d}{d-1}\sum_i P_{i,i}^2-\frac{1}{d-1}}\;,\nonumber\\
\end{eqnarray}
where $\sum_\pi$ denotes the sum over all possible permutations of
$P_{i,j}$. The first line admits for a multi-slit scenario the
following simple interpretation: it is the difference of the
probability that the particle transverses the slit, e.g. $P_{0,0}$,
minus the probabilities that the particle takes the way through all
the other slits weighted by $d-1$, summed over all slits. For $d=2$
it is clearly equivalent to the prior definition of predictability,
it ranges from zero to one and is equal to one if one has hundred
per cent information about a possible measurement outcome and it is
equal to zero if one does not have any information about which
degree of freedom would most likely to be measured. Thus is meets
all of our conceptual requirements.

%It is applicable to a system of an arbitrary dimension d and has the
%following properties:
%\begin{itemize}
%    \item For $d=2$ it is equivalent to the prior definition of predictability
%    \item It ranges from zero to one
%    \item It is equal to one if one has hundred per cent information about a possible measurement outcome
%    \item It is equal to zero if one does not have any information about which degree of freedom would most likely to be measured
%\end{itemize}
Of course it is only one of many ways to describe multi-level
predictability, but one that meets all of our conceptual
requirements. A different one was introduced in Ref.~\cite{Janos2}.

The coherence is easier to define, it is more straightforward. We
can just take the sum over all two dimensional coherences:
\begin{eqnarray}
C_{coh,g}(\rho)&:=&\left(\frac{2d}{d-1}\sum_{j=1}^{d-1}\sum_{i<j}\left|P_{i,j}\right|^2\right)^{\frac{1}{2}}\nonumber\\
&=&\left(\frac{d}{d-1}\left(Tr(\rho^2)-\sum_i
P_{i,i}^2\right)\right)^{\frac{1}{2}}
\end{eqnarray}
The last equation is obtained by using
$Tr(\rho^2)=\sum|P_{i,j}|^2=\sum P_{i,i}^2+2\sum_{i,j,i<j}
|P_{i,j}|^2$. Again it meets also the conceptual requirements as for
$d=2$ it is equivalent to the prior definition of coherence, it
ranges from zero to one and is equal to one if one has the most
coherent superposition of all degrees of freedom and it is equal to
zero if one has hundred per cent information about a possible
measurement outcome.

Let us now consider the sum of generalized predictability and
coherence
\begin{eqnarray}\label{complddim}
&&P_g^2(\rho)+C_{coh,g}(\rho)^2=\frac{d}{d-1}Tr(\rho^2)-\frac{1}{d-1}=-M^2+1\nonumber\\
&&\qquad\Longrightarrow
\underbrace{P_g^2(\rho)+C_{coh,g}(\rho)^2}_{{\cal
S}_g^2(\rho)}+M^2(\rho)=1
\end{eqnarray}
The last equation is the generalized Bohr complementary relation for
$d$--dimensional systems we searched for and helps to understand
entanglement in multiqudit systems.

\section{The entanglement measure and its bounds} Let us now proceed
to entangled systems. For pure states it is well known that
entanglement can be solely quantified by considering the mixedness
or entropy of the subsystems. For a $n$--partite system where each
subsystem $s$ is given by the reduced density matrix
$\rho_s=Tr_{[1,\dots,s-1,s+1,\dots,n]}(|\psi\rangle\langle\psi|)$ of
dimensionality $d_s$, the sum of the mixednesses defines an
entanglement measure
\begin{eqnarray}
E(|\psi\rangle):=\sum_{s=1}^n M^2(\rho_s):=\sum_{s=1}^n
M_s^2(\rho)\,.
\end{eqnarray}
This is an entanglement measure, i.e. non-increasing under LOCC
(local operations and classical communications), additive and
meeting all requirements to be an entanglement monotone (see e.g.
Ref.~\cite{buchleitnermulti}). The same is true for the Von Neumann
entropies of the subsystems defining entanglement of formation
\cite{Benn}, but the linear entropy $M^2$ bears the benefit that it
can be operationally obtained as we will show.

\subsection{Example: A tripartite qubit state}

For convenience and clarity we consider now a tripartite qubit
system and generalize then for the multipartite case. Consider the
tripartite qubit state
\begin{eqnarray}
|\psi\rangle=\sum_{i,j,k=0}^1 a_{ijk}|ijk\rangle\;,
\end{eqnarray}
then the mixednesses of the subsystem, e.g. of the first qubit, is
straightforward calculated and equivalent to the following
expressions obtained by simple algebra: \begin{widetext}
\begin{eqnarray*}
\lefteqn{M_1^2(\psi)=M^2(\rho_1)=}\nonumber\\
&&\sum_{k=0}^1
\sum_{i_1\not=i'_1;i_2\not=i'_2}\bigl|\langle\psi|(\sigma\otimes\sigma\otimes\mathbbm{1})
\left(|i_1\, i_2\, k\rangle\langle i_1\, i_2\, k|-|i'_1\, i'_2\,
k\rangle\langle
i'_1\, i'_2\, k|\right)|\psi^*\rangle\bigr|^2\nonumber\\
 &+&\sum_{k=0}^1
\sum_{i_1\not=i'_1;i_3\not=i'_3}\bigl|\langle\psi|(\sigma\otimes\mathbbm{1}\otimes\sigma)
\left(|i_1\, k\, i_3\rangle\langle i_1\, k\, i_3|-|i'_1\, k
\,i'_3\rangle\langle i'_1\, k\, i'_3|\right)|\psi^*\rangle\bigr|^2\nonumber\\
 &+& \sum_{i_1\not=i'_1;i_2\not=i'_2;i_3\not=i'_3}
\bigl|\langle\psi|(\sigma\otimes\sigma\otimes\sigma) \left(|i_1\,
i_2\, i_3\rangle\langle i_1\, i_2\, i_3|-|i'_1\, i'_2
\,i'_3\rangle\langle i'_1\, i'_2\,i'_3|\right)|\psi^*\rangle\bigr|^2%\nonumber\\
\end{eqnarray*}
\end{widetext}
where $\sigma$ is the flip operator, i.e. here the Pauli matrix
$\sigma_x$. One sees that the squared mixedness of one subsystem is
obtained by flipping once in each other subsystem and flipping in
all subsystems. Straightforward one obtains the squared mixednesses
of the two other systems. As the sum of all mixednesses, $E(\psi)$,
is an entanglement measure it is obvious to add all terms which
contain two flips to one quantity which we denote by
$(\textbf{C}_{(2)})^2$ and all terms with three flips to
$(\textbf{C}_{(3)})^2$. Thus we have separated the total entanglement,
$E(\psi)$, into a sum of terms containing two or three flips, which
we name in the following as $m$--flip concurrence.

\subsection{The $m$--concurrence for pure states}

For multipartite qudit systems the same works. The flip operators
can be defined in the following way for a qudit system of dimension
$d$:
\begin{eqnarray}
&&\sigma_{kl}^{d\times
d}|k\rangle\;=\;|l\rangle,\qquad\sigma_{kl}^{d\times
d}|l\rangle\;=\;|k\rangle\quad\textrm{and}\nonumber\\
&&\sigma_{kl}^{d\times d}|t\rangle\;=\;0\quad \forall\; t\neq k,l
\end{eqnarray}
with $k,l\in\left\{0,1,\dots,d-1\right\}$. Note that this is just another definition of the d-dimensional symmetric Gellmann matrices. The squared $m$--flip
entanglement is given by the sum over all possible permutations of $m$
flips, i.e. for $n$ systems there are
$\left(\begin{array}{c}n\\m\end{array}\right)$ possibilities and
each system where a flip is performed is denoted by $\alpha_j$ and
in order to avoid multiple counting we order the set:
$\{\alpha_j\}:=\alpha_1,\alpha_2,\dots,\alpha_m$, where
$\alpha_1<\alpha_2<\dots<\alpha_m$:
\begin{eqnarray}
(\textbf{C}_{(m)})^2&=&\sum_{\{\alpha_j\}}\textbf{C}^2_{\{\alpha_j\}}\;.
\end{eqnarray}
Each possibility of $m$ flips is derived by
\begin{eqnarray*}
\textbf{C}^2_{\{\alpha_j\}}:=\sum_{set}\left|\langle\psi|
\hat{O}_{\{\alpha_j\}}(|\{i_n\}\rangle\langle\{i_n\}|-|\{i'_n\}\rangle\langle\{i'_n\}|)|\psi^*\rangle\right|^2%\nonumber\\
\end{eqnarray*}
where
\begin{eqnarray}
\sum_{set}:=\sum_{i\in\{\alpha_j\}}\sum_{l_i=1}^{d_i-1}\sum_{k_i<l_i}\underbrace
{\sum_{\{i_n\}\neq\{i'_n\}}}_{i_t=i'_t \forall t\notin\{\alpha_j\}}
\end{eqnarray}
and
\begin{eqnarray}
\hat{O}_{\{\alpha_j\}}:=\left(\sigma_{k_il_i}^{s\in\{\alpha_j\}},\mathbbm{1}^{s\notin\{\alpha_j\}}\right)
\end{eqnarray}
which defines a $n$--tensor product where the flip operator are
positioned at $\alpha_j$ and else the unity is taken. Note that this
works for n--partite systems with dimensions $d_1$ to $d_n$. And the
total entanglement is given by
\begin{eqnarray}
E(\psi):=\sum_{m=2}^n\textbf{C}^2_{(m)}=\sum_{s=1}^n
M_s^2(|\psi\rangle\langle\psi|)
\end{eqnarray}
This, for pure systems, is equal to the sum of the squared
mixednesses of the subsystems. Also if we set $d=n=2$ this is just
the definition of Wootter's concurrence  \cite{Wootters} multiplied by two.\\

\subsection{The $m$--concurrence for mixed states}

Pure states are quite a strong restriction and for mixed states, the
mixedness of the subsystem will not suffice, because it stems from
both entanglement and classical uncertainty. The $m$--flip
concurrence for mixed density matrices can be defined by:
\begin{eqnarray}
(\textbf{C}^m_g(\rho))^2:=\textbf{inf}_{|\psi_i\rangle,p_i}\sum_{|\psi_i\rangle,p_i}p_i\;(\textbf{C}_{(m)}(|\psi_i\rangle))^2
\end{eqnarray}
That this is part of an analytically correct entanglement measure is
obvious, because:
\begin{itemize}
    \item[(1)] $\sum_{m=2}^n(\textbf{C}_{(m)}(|\psi\rangle))^2$ is an entanglement measure for pure
    states.
    \item[(2)] Any separable density matrix can be decomposed into a convex sum of pure separable states, hence the infimum
     equals zero for all separable states.
    \item[(3)] Any entangled density matrix's decomposition contains at least one entangled pure state, hence $\textbf{C}^m_g(\rho)$
     is part of an entanglement measure which is nonzero for all entangled states.
\end{itemize}

\subsection{Bounds on the $m$--flip concurrence} We can derive
bounds for the $m$--flip concurrence by defining in an analogous way
to Hill and Wootters flip density matrix \cite{Wootters} the
$m$--flip density matrix:
\begin{eqnarray}
\widetilde{\rho}_{\left\{\alpha_j\right\}}^m=O_{\left\{\alpha_j\right\}}(|\{i_n\}\rangle
\langle\{i_n\}|-|\{i'_n\}\rangle\langle\{i'_n\}|)\;\rho^*
\cdot\nonumber\\ \cdot\; O_{\left\{\alpha_j\right\}}
(|\{i_n\}\rangle\langle\{i_n\}|-|\{i'_n\}\rangle\langle\{i'_n\}|)
\end{eqnarray}
and calculating the $\lambda_m^{\left\{\alpha_j\right\}}$'s which
are the squared roots of the eigenvalues of
$\widetilde{\rho}_{\left\{\alpha_j\right\}}^m \rho$. The bound can
be derived as (analogously to Ref.~\cite{buchleitner1})
%\begin{widetext}
\begin{eqnarray}
\lefteqn{B^m(\rho):=}\nonumber\\
&&\left(\sum_{\{\alpha_j\}}\sum_{set}\textbf{max}\left[0,2\textbf{max}
\left(\{\lambda_m^{\left\{\alpha_j\right\}}\}\right)-\sum
\{\lambda_m^{\left\{\alpha_j\right\}}\}\right]^2\right)^{\frac{1}{2}}\nonumber
\end{eqnarray}
%\end{widetext}
This is not equivalent to the convex roof, but a good boundary %for $\textbf{C}^m_g(\rho)$:
$B^m(\rho)\leq \textbf{C}^m_g(\rho)$ as it agreed in all cases we
tried with the criterion of partial positive transposition. In literature there
exist several possible suggestions how to improve
these bounds \cite{buchleitner1}.

%\textbf{The $m$--flip entanglement for multi qubit systems:}
\section{$m$--flip concurrence and $m$--partite entanglement}

Does the $m$--flip concurrence, $\textbf{C}_{(m)}$, also describe the
desired $m$--partite entanglement? The answer is no, a simple
counter example e.g. for tripartite qubits is
\begin{eqnarray}\label{halvesep}\frac{1}{\sqrt{2}}\lbrace
|00\rangle+|11\rangle\rbrace\otimes(\cos\alpha|0\rangle+\sin\alpha|1\rangle)\;.\end{eqnarray}
Here only the first and second particle are entangled the third one
not, thus the tripartite entanglement should be zero, but the
$3$--flip entanglement derives to:
\begin{eqnarray}
\textbf{C}^2_{(3)}&=& 4|\cos\alpha \sin\alpha|^2\;.
\end{eqnarray}
Moreover, the $m$--flip concurrence is not invariant under local
unitaries. However, if we introduce ``corrections'' to the $m$--flip
concurrence such that these new quantities are invariant under local
unitaries we obtain the desired $m$--partite entanglement. For sake of
simplicity we stick here to the case of three qubits.

\begin{figure}
\includegraphics[width=\columnwidth]{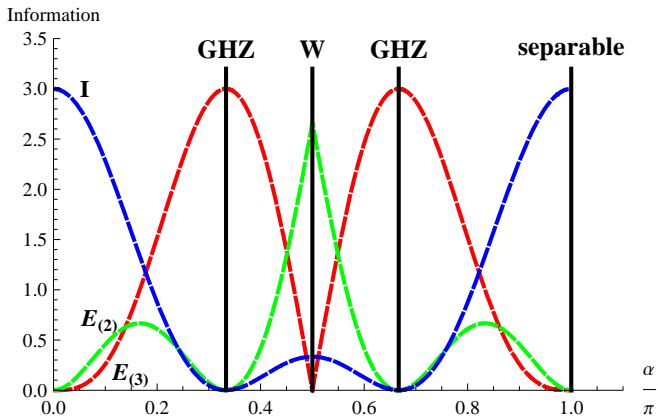}
\caption{Here the information content in bits of the state
$\rho_\alpha$, Eq.~(\ref{phialpha}), is plotted. The colored,
thickened and dashed curves are the single properties
$I=\sum_{s=1}^3 S_s$ (blue),
 the $2$--partite entanglement
$E_{(2)}=E_{12}+E_{13}+E_{23}$ (green) and the $3$--partite
entanglement $E_{(3)}=E_{123}$ (red). For the GHZ--state
$\alpha=\frac{\pi}{3}/\frac{2\pi}{3}$ the $3$--partite entanglement
is maximal while for the W--state $\alpha=\frac{\pi}{2}$ the
$2$--partite entanglement is maximal. For $\alpha=\frac{\pi}{6}\
\frac{5\pi}{6}$ we obtain another interesting state, the
$2$--partite entanglement has a second
maximum.}\label{fig:3qubitstates}
\end{figure}

\textbf{Three qubits states:} There exist two entangled states, the
well known GHZ--state and the W--state, which are obviously in a
physically different way entangled. If one traces over one
subsystem, in first case one gets a separable state and in the
second case an entangled state.

For that let us discuss the states
$\rho_{\alpha}=|\phi(\alpha)\rangle\langle\phi(\alpha)|$ with
\begin{eqnarray}\label{phialpha}
|\phi(\alpha)\rangle&=&\frac{\sin{\alpha}}{\sqrt{3}}\{
|001\rangle+|010\rangle+|100\rangle\}+\cos\alpha\;|111\rangle\;,\nonumber\\
\end{eqnarray}
which are superposition of the W--state and a separable state. It is
also plotted in Fig.~\ref{fig:3qubitstates}. For
$\alpha=\frac{\pi}{3}/\frac{2\pi}{3}$ the state is unitary
equivalent to the GHZ--state
($\frac{1}{\sqrt{2}}\lbrace|000\rangle+|111\rangle\rbrace$). This is
at the first side surprising, because the $3$--flip and the
$2$--flip concurrences derive to
\begin{eqnarray}
\textbf{C}^2_{(3)}&=&0\nonumber\\
\textbf{C}^2_{(2)}&=&\frac{8}{3}(\sin^4(\alpha)+3
\cos^2(\alpha)\sin^2(\alpha))\;,
\end{eqnarray}
i.e. the $3$-flip is zero. For the W--state ($\alpha=\frac{\pi}{2}$)
on the other side the $2$--flip has a local minimum. These facts
suggest that we have to ``correct'' the flip concurrences to obtain
the $m$--partite entanglement in the following way
\begin{eqnarray}
E_{(3)}&=&E_{123}\;=\;\textbf{C}^2_{(3)}(\rho_\alpha)+\textbf{C}^2_{(2)}(\rho_\alpha)\nonumber\\
\lefteqn{-\lbrace\textbf{C}^2_{(2)}\left(Tr_1(\rho_\alpha)\right)+
\textbf{C}^2_{(2)}(Tr_2(\rho_\alpha))+\textbf{C}^2_{(2)}(Tr_3(\rho_\alpha))\rbrace}\nonumber\\
E_{(2)}&=&E_{12}+E_{13}+E_{23}\nonumber\\
&=&\textbf{C}^2_{(2)}(Tr_1(\rho_\alpha))+
\textbf{C}^2_{(2)}(Tr_2(\rho_\alpha))+\textbf{C}^2_{(2)}(Tr_3(\rho_\alpha))\;.\nonumber\\
\end{eqnarray}
It is clear that the sum is unchanged, because we add the sum of the
$2$--flip concurrence of the reduced density matrices and subtract it.
As the $2$--flip concurrence of
the subsystems are here equivalent to two times the Wootters-Hill
concurrence \cite{Wootters} this sum is clearly invariant under
local unitaries. Therefore all that is left to show is that $E_{(3)}$
cannot get negative. This is easily proven because the entanglement
stored in the subsystems can only be lower or at most equivalent to
the entanglement stored in the subsystems. This can of course be generalized for mixed
states through the convex roof.

With these definitions of the $2$-- and $3$--partite entanglement we
obtain a simple interpretation of the physical difference of the W--
and the GHZ--entanglement. The state $\rho_\alpha$,
Eq.~(\ref{phialpha}), is for $\alpha=0$ separable, the single
property in each subsystem is maximal (see also
Fig.~\ref{fig:3qubitstates}). As $\alpha$ increases the single
properties $I(\rho_\alpha)=\sum_{s=1}^3 S_s^2$ have to decrease due
to Bohr's complementary relation ($S_s^2+M_s^2=1$) as their
mixednesses increase. For $\alpha=\frac{\pi}{3}$ or
$\alpha=\frac{2\pi}{3}$ the single properties are zero, thus
entanglement is maximal and it is a genuine $3$--partite entangled
state, the GHZ--state. For $\alpha=\frac{\pi}{2}$ the $2$--partite
entanglement is maximal, while the single properties obtain a local
maximal and the $3$--partite entanglement is zero as desired. There
is another interesting state e.g. at
$\alpha=\frac{\pi}{6}/\frac{5\pi}{6}$, here $2$--partite
entanglement has another maximum (see Fig.~\ref{fig:3qubitstates}).

\begin{figure}
\includegraphics[width=\columnwidth]{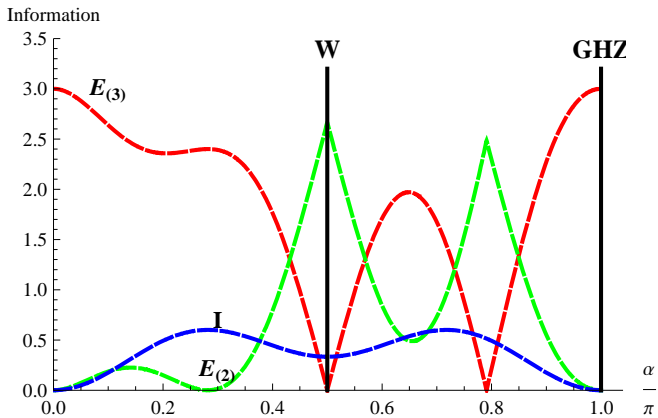}
\caption{Here the information content in bits of the state
$\tau_\alpha$, Eq.~(\ref{taualpha}), is plotted. The colored,
thickened and dashed curves are the single properties
$I=\sum_{s=1}^3 S_s$ (blue),
 the $2$--partite entanglement
$E_{(2)}=E_{12}+E_{13}+E_{23}$ (green) and the $3$--partite
entanglement $E_{(3)}=E_{123}$ (red). The single properties $I$ are
symmetric, however, the genuine $2$-- and $3$--partite entanglement
are not.}\label{fig:ghzwsup}
\end{figure}

Also the above counter example of a bi--separable state,
Eq.~(\ref{halvesep}), obtains the desired interpretation as the
$3$--partite entanglement $E_{(3)}$ derives to zero and the
$2$--partite entanglement to $E_{(2)}=E_{12}+E_{13}+E_{23}=2+0+0$,
see also Fig.~\ref{fig:anystate}~(d).

In Fig.~\ref{fig:ghzwsup} we show the superposition of W and GHZ: $\tau_\alpha=|\psi(\alpha)\rangle\langle\psi(\alpha)|$
\begin{eqnarray}\label{taualpha}
|\psi(\alpha)\rangle&=&\frac{\sin{\alpha}}{\sqrt{3}}\bigl\lbrace
|001\rangle+|010\rangle+|100\rangle\bigr\rbrace\nonumber\\
&&+\frac{\cos{\alpha}}{\sqrt{2}}\bigl\lbrace|111\rangle+|000\rangle\bigr\rbrace\;.
\end{eqnarray}
The single properties are symmetric, however, the tripartite and the
bipartite entanglement depend on the particular superposition. One
finds for example another interesting state at ($\alpha=\simeq 0.8
\pi$) which maximizes the bipartite entanglement $E_{(2)}$ while the
tripartite entanglement $E_{(3)}$ is zero (see
Fig.~\ref{fig:ghzwsup}).

\begin{figure}
\includegraphics[width=\columnwidth]{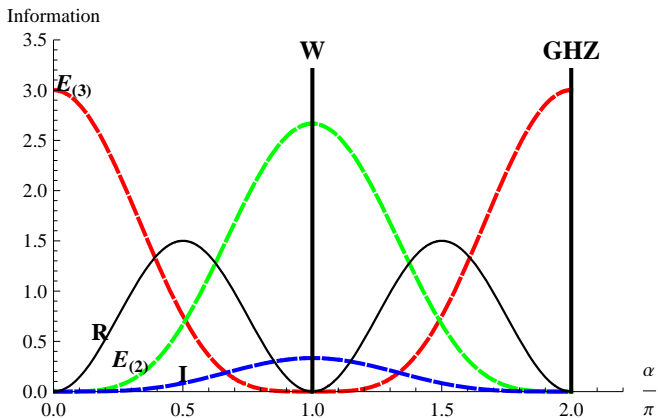}
\caption{Here the information content in bits of the state
$\sigma(\alpha)$, Eq.~(\ref{mixedsigma}), is plotted. The colored,
thickened and dashed curves are the single properties
$I=\sum_{s=1}^3 S_s$ (blue),
 the $2$--partite entanglement
$E_{(2)}=E_{12}+E_{13}+E_{23}$ (green) and the $3$--partite
entanglement $E_{(3)}=E_{123}$ (red). The thin, not dashed curve is
$R(\rho)$, which is the lack of information about the state.
}\label{fig:ghzwmix}
\end{figure}

For mixed states, if the optimal bounds are known, the
$m$--entanglement derives in the very same way. In
Fig.~\ref{fig:ghzwmix} we have chosen a mixture of the GHZ--state and
the W--state, i.e.
\begin{eqnarray}\label{mixedsigma}
\sigma(\alpha)&=&\sin^2(\frac{\alpha}{2})\;\rho_{W}+\cos^2(\frac{\alpha}{2})\;\rho_{GHZ}\;
\end{eqnarray}
The complementing missing information, due to classical lack of
knowledge about the quantum state, $R(\sigma(\alpha))$, is maximal
for $\alpha=\frac{\pi}{2}$ as expected.

To sum up for $3$ qubits we have shown that in a simple and evident
way the $2$--flip and $3$--flip concurrences can be made invariant
under local unitaries and these new quantities, the $2$-- and
$3$--partite entanglement, capture then the desired physical
differences e.g. of the GHZ--state, the W--state, the bi--separable
state and of mixed states.

\section{The information content of n--partite multidimensional systems}

Now returning to the first equation, Eq.~(\ref{voll}), we see the
separation of information in multi-qudit systems: $I(\rho)$
quantifies all locally obtainable information as it is just the sum
over all obtainable information in every subsystem. Every physical
system with d-degrees of freedom can carry one dit of information,
which can be separated into predictability and coherence. Of course
the combination of n-systems can carry n-dits of information, just
as in classical systems. The main difference then is that they
cannot all be locally obtained, as for pure entangled systems the
sum over all locally obtainable information does not yield n. Here
the information is encoded in entanglement, which in itself seems to
be separable into different classes of entanglement, as we have seen
for the three qubit example. Note that the local information
$I(\rho)$ is always additive and for the entanglement $E(\rho)$
strict additivity is only proven for pure states. For mixed systems
only subadditivity of $E(\rho)$ can be proven (in the very same way
as with entanglement of formation), but additivity is strongly
expected (in this case $R(\rho)$ can clearly only contain classical
uncertainty).

\section{Conclusion}

We started from Bohr's complementarity relation, which we
generalized for $d$--dimensional systems, yielding the information
content in every subsystem of a multipartite state $\rho$ through
the single property ${\cal S}_s(\rho)$. From this relation we are
able to consistently quantify the information content in
entanglement of pure states as the missing information needed to
complement the single properties of all subsystems as
$n-\sum_{s=1}^n{\cal S}^2_s(\rho)=\sum_{s=1}^n M^2_s(\rho)$. Then we
showed that this information can be operationally obtained, thus
opening the possibility to derive bounds for the entanglement of
multipartite multidimensional systems. Moreover, we have shown that
the operations necessary to obtain this measure can be separated
into different classes of concurrences, i.e. characterized by the
amount of flips $m$, the $m$--flip concurrences. Finally that these
can be modified to meet our understanding of multipartite
entanglement as explicitly shown for three qubits one can correct
the $m$--flip concurrences to $m$--partite entanglement. This
explains the different kind of entanglement of the GHZ--state and
the W--state, summarized in Fig.~\ref{fig:anystate} and in
Fig.~\ref{fig:3qubitstates}. Moreover, it gives the correct
entanglement of any bi--separable state. Last but not least we
presented a mixed state, where the single properties are
complemented by the classical lack of knowledge about the quantum
state under investigation. In what way this generalizes for
multipartite systems consistently is left for further investigation.

In summary, we found a multidimensional, multipartite and
operational entanglement measure which admits a separation into
different classes of entanglement and herewith we obtain the
information content of any discrete state. This knowledge obviously
helps to e.g. decide given a certain state quantum communication or
quantum cryptography is possible and to what extend.

\begin{small}
\textbf{Acknowledgement:} Thanks to R.A. Bertlmann and Ph. Krammer
for enlightening discussions.
\end{small}

\end{document}